\documentclass[
 aip,
 amsmath,amssymb,
 reprint,
]{revtex4-1}

\usepackage{graphicx}
\usepackage{dcolumn}
\usepackage{bm}

\usepackage[utf8]{inputenc}
\usepackage[T1]{fontenc}
\usepackage{mathptmx}
\usepackage{etoolbox}
\usepackage{xcolor}
\usepackage[normalem]{ulem}

\makeatletter
\def\@email#1#2{%
 \endgroup
 \patchcmd{\titleblock@produce}
  {\frontmatter@RRAPformat}
  {\frontmatter@RRAPformat{\produce@RRAP{*#1\href{mailto:#2}{#2}}}\frontmatter@RRAPformat}
  {}{}
}
\makeatother

\begin{document}

\preprint{AIP/123-QED}

\title{Device variability of Josephson junctions induced by interface roughness}
\author{Yu Zhu}
\email{eric@nanoacademic.com}
\affiliation{Nanoacademic Technologies Inc., Suite 802, 666 rue Sherbrooke Ouest, Montréal, Québec H3A 1E7, Canada}

\author{F\'elix Beaudoin}
\affiliation{Nanoacademic Technologies Inc., Suite 802, 666 rue Sherbrooke Ouest, Montréal, Québec H3A 1E7, Canada}

\author{Hong Guo}
\affiliation{Department of Physics, McGill University, Montréal, Québec, H3A 2T8, Canada}

\date{\today}

\begin{abstract}
As quantum processors scale to large qubit numbers, device-to-device
variability emerges as a critical challenge. Superconducting qubits are
commonly realized using Al/AlO$_{\text{x}}$/Al Josephson junctions operating
in the tunneling regime, where even minor variations in device geometry can
lead to substantial performance fluctuations. In this work, we develop a
quantitative model for the variability of the Josephson energy $E_{J}$
induced by interface roughness at the Al/AlO$_{\text{x}}$ interfaces. The
roughness is modeled as a Gaussian random field characterized by two
parameters: the root-mean-square roughness amplitude $\sigma $ and the
transverse correlation length $\xi $. These parameters are extracted from
the literature and molecular dynamics simulations. Quantum transport is
treated using the Ambegaokar--Baratoff relation combined with a local
thickness approximation. Numerical simulations over $5,000$ Josephson
junctions show that $E_{J}$ follows a log-normal distribution. The mean
value of $E_{J}$ increases with $\sigma $ and decreases slightly with $\xi $,
while the variance of $E_{J}$ increases with both $\sigma $ and $\xi $. These results paint a quantitative and intuitive picture of Josephson energy variability induced by surface roughness, with direct relevance for junction design.
\end{abstract}
\maketitle


\section{Introduction}

Superconducting circuits are among the most promising platforms for
scalable quantum computing, combining lithographic
manufacturability with strong, low-loss nonlinearity from the Josephson effect \cite%
{ref01-outlook}. 
At the heart of widely used transmon-style devices is an
Al/AlO$_{\text{x}}$/Al Josephson junction shunted by a capacitance, forming
a weakly anharmonic quantum oscillator whose lowest two eigenstates $%
\left\vert 0\right\rangle $ and $\left\vert 1\right\rangle $ serve as the
computational basis \cite{ref02-qubit-design}. In this picture, the qubit
transition frequency is set primarily by the Josephson energy $E_{J}$ and
charging energy $E_{C}$, so any process-dependent change in the junction's
critical current (and hence $E_{J}$) directly shifts the qubit spectrum and
operating point. State-of-the-art superconducting quantum processors now
contain on the order of $10^{2}$ to $10^{3}$ qubits, exemplified by IBM's $%
1,121$-qubit \textquotedblleft Condor\textquotedblright\ superconducting
processor \cite{ref03-Condor}. However, fault-tolerant quantum computation
for societally relevant tasks remains far beyond this scale. For example,
detailed resource estimates for Shor's algorithm indicate that factoring a
2048-bit RSA integer would require on the order of tens of millions of noisy
physical qubits in a fast-runtime scenario (e.g., $\sim 20$ million noisy
qubits for an 8-hour attack under surface-code assumptions) \cite%
{ref04-RSA2048}. This large gap motivates careful attention to device yield
and uniformity as superconducting platforms scale toward much larger systems.

Josephson junction variability becomes a central challenge at scale because
junctions operate in the tunneling regime, where the critical current $I_{c}$
depends exponentially on microscopic barrier details; small
perturbations---such as local Al/AlO$_{\text{x}}$ thickness fluctuations,
effective barrier-height variations, or contamination atoms---can produce
significant shifts in $E_{J}$, qubit frequency, and gate fidelity.
Experimentally, substantial efforts over the past two decades have focused
on improving junction reproducibility and understanding microscopic sources
of variation, including wafer- and chip-scale resistance/critical-current
uniformity studies \cite{ref05-WaferScaleJJ,ref06-JJ-reprod1},
process-optimization of shadow evaporation/oxidation \cite%
{ref06-JJ-reprod1,ref07-JJ-reprod2}, mitigation strategies for
critical-current fluctuations \cite{ref08-mitigation}, alternating-bias assisted annealing\cite{pappas2024alternating}, 
and direct microscopy
of Al/AlO$_{\text{x}}$ barrier microstructure and thickness distributions 
\cite{Zeng2015,Fritz2019}. In this work, we focus on variability in $%
E_{J} $ induced by Al/AlO$_{\text{x}}$ interface roughness, and present a
quantum-mechanical modeling framework that combines quantum transport theory
with Gaussian random field models and molecular dynamics simulations to
connect microstructural disorder to junction-level Josephson energy
statistics.

\section{Interface roughness model}

The rough interface between the aluminum (Al) leads and the aluminum oxide
(AlO$_{\text{x}}$) layer is modeled as a Gaussian random field with a
specialized power sepctral density. Denote the interface height by $h\left(
x,y\right) $, where $x$ and $y$ are the tranverse coordinates. It is assumed
that 
\begin{equation}
\left\langle h\left( x,y\right) \right\rangle =0,  \label{eq05}
\end{equation}%
and%
\begin{equation}
\left\langle h\left( x,y\right) h\left( 0,0\right) \right\rangle =\sigma
^{2}e^{-\frac{x^{2}+y^{2}}{\xi ^{2}}},  \label{eq06}
\end{equation}%
where $\sigma $ is the root-mean-square (RMS) of the interface height along
the transport direction, and $\xi $ is the transverse correlation length.
Given $\sigma $ and $\xi $, a rough interface can be generated by filtering
white noise in Fourier space using the corresponding power spectral density,
followed by an inverse Fourier transform. Two generated rough interfaces are
shown in Fig. \ref{fig04} (a) and (b).

The transverse correlation length $\xi $ is set by the characteristic
lateral length scale of microstructural features in the Al leads---most
notably the Al grain size that imprint thickness variations onto the AlO$_{%
\text{x}}$ barrier. In Al/AlO$_{\text{x}}$/Al tunnel junction stacks
characterized by TEM, the Al electrodes are commonly observed to be
polycrystalline with columnar grains on the order of a few tens of
nanometers. For example, Liu \textit{et al}. report crystalline columnar
grains of roughly $\sim $ 20 nm $\times $ 30 nm in qubit Josephson junction
electrodes \cite{Fritz2019}. A systematic TEM study by Nik \textit{et al.}
further shows that the average lateral grain size depends strongly on the Al
film thickness: for Al films of 15 nm and 60 nm thickness on oxidized Si,
the mean grain size is reported to be on the order of $\sim $ 38 nm and $%
\sim $ 92 nm, respectively, illustrating that thicker Al films tend to
develop larger grains \cite{ref22-Al02}. Under other deposition conditions,
Fritz \textit{et al.} report that the lower Al layer can exhibit a broad
grain size distribution, including grains below $\sim $ 50 nm, many grains
in the $\sim $ 50 nm-150 range, and grains extending up to hundreds of
nanometers ($\sim $ 900 nm) and even $\sim $ 1 $\mu $m in lateral size \cite%
{Fritz2019}. These experimental observations suggests choosing $\xi $ to be
comparable to the typical lateral grain size for the specific Al growth
recipe used in the junction fabrication.

The RMS height $\sigma $ is set by the vertical amplitude of
thickness/height fluctuations, primarily controlled by the Al surface
roughness and oxidation/growth nonuniformity. Zeng \textit{et al.} directly
measured the local AlO$_{\text{x}}$ thickness in Al/AlO$_{\text{x}}$/Al
junctions using atomic-resolution STEM and fit the thickness histogram with
a Gaussian. They report mean thicknesses 1.66--1.88 nm and standard
deviations 0.326--0.372 nm in three samples \cite{Zeng2015}. Fritz \textit{%
et al. } report AlO$_{\text{x}}$ thickness variation values
ranging from 0.29 nm down to 0.11 nm depending on growth conditions \cite{Fritz2019}. These experiments suggest a representative range of $\sigma $ $\sim $ 0.1--0.3 nm
for high-quality AlO$_{\text{x}}$ tunnel barriers.

\begin{figure}
	\centering
	\includegraphics[width=8.7cm]{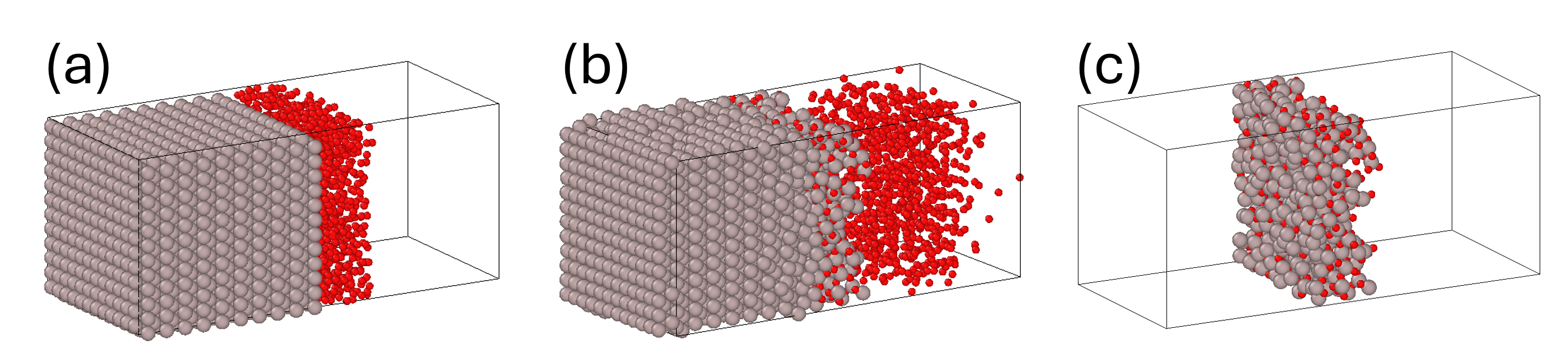}
	\caption{MD simulation of AlO$_{\text{x}}$ growth. (a) At $t=0$, fcc(100) Al surface and O$_{\text{2}}$ molecules. (b) At $t=3$ ps, oxidized Al surface and O$_{\text{2}}$ molecules. (c) AlO$_{\text{x}}$ layer extracted from (b).
    }
	\label{fig03}
\end{figure}

Since $\sigma $ is much smaller than $\xi $ and more difficult to
characterize experimentally, we also perform molecular dynamics (MD)
simulations to estimate $\sigma $. The initial simulation cell contains an
Al surface, O$_{\text{2}}$ molecules, and a vacuum region. In total, the
system has $4,380$\ atoms, comprising $2,880$ Al atoms and $750$ O$_{\text{2}%
}$ molecules, as illustrated in Fig. \ref{fig03} (a). The Al surface is modeled as an fcc(100) slab consisting of 20 atomic layers, with 144 atoms per layer and a 40 Å vacuum region. 
The O$_{\text{2}}$ molecules are placed at random positions with random
orientations in a $10$ \AA\ thick region between the Al surface and the
vacuum. MD simulations are carried out in the NVT ensemble at a fixed
temperature $300$ K and a fixed volume $34.17$ \AA\ $\times $ $34.17$ \AA\ $%
\times $ $78.26$ \AA , with periodic boundary conditions in the transverse
directions. Interatomic forces and energies are computed using CHGNet, a
charge-informed graph-neural-network machine learning interatomic potential
pretrained on the Materials Project Trajectory dataset \cite{ref24-chgnet}.

MD simulations are performed for $3$ ps with a time step of $1$ fs. During
the first $\sim $2 ps, O$_{\text{2}}$ molecules react with the Al surface
and form an amorphous AlO$_{\text{x}}$ layer with a thickness about 1 nm, as
shown in Fig. (b). Then the oxide growth saturates and the remaining O$_{%
\text{2}}$ molecules diffuse into the vacuum region. After generating the
Al/AlO$_{\text{x}}$ interface, we remove Al lead atoms and O$_{\text{2}}$
molecules from the system to expose the AlO$_{\text{x}}$ surface, as shown
in Fig. \ref{fig03} (c). To extract a smooth height profile, we employ a
rolling-ball procedure: a probe sphere of radius $R=3$ \AA\ is rolled over
the AlO$_{\text{x}}$ surface on the Al side to define a continuous surface
envelope. The RMS\ of the surface $\sigma $ is obtained as $0.97$ \AA , $%
0.76 $ \AA , $0.81$ \AA\ in three independent MD simulations. We note that $%
\sigma $ values extracted from our MD simulations are smaller than the
experimentally reported values. This discrepancy is likely due to the
idealized geometry used in the simulations: the Al surfaces are assumed to
be atomically flat, whereas real devices typically exhibit additional Al
surface roughness that increases the overall height fluctuations.

\section{Quantum transport model}

In a Josephson junction, the Josephson energy $E_{J}$ is related to the
critical current $I_{c}$ by%
\begin{equation}
E_{J}=\frac{\hbar }{2e}I_{c},  \label{eq01}
\end{equation}%
where $\hbar $ is is the reduced Planck constant and $e$ is the elementary
charge. The critical current can be calculated by solving the Bogoliubov--de
Gennes (BdG) equation%
\begin{equation}
\left( 
\begin{array}{cc}
H_{0}\left( \mathbf{r}\right) -E_{F} & \Delta \left( \mathbf{r}\right) \\ 
\Delta ^{\ast }\left( \mathbf{r}\right) & -\left( H_{0}\left( \mathbf{r}%
\right) -E_{F}\right)%
\end{array}%
\right) \left( 
\begin{array}{c}
u\left( \mathbf{r}\right) \\ 
v\left( \mathbf{r}\right)%
\end{array}%
\right) =E\left( 
\begin{array}{c}
u\left( \mathbf{r}\right) \\ 
v\left( \mathbf{r}\right)%
\end{array}%
\right) ,  \label{eq02}
\end{equation}%
where%
\[
H_{0}\left( \mathbf{r}\right) =-\frac{\hbar ^{2}}{2m}\nabla ^{2}+V\left( 
\mathbf{r}\right) 
\]%
is the single-particle Hamiltonian, $E_{F}$ is the Fermi energy, $\Delta
\left( \mathbf{r}\right) $ is the superconducting pair potential, $u\left( 
\mathbf{r}\right) $ and $v\left( \mathbf{r}\right) $ are the electron- and
hole-like components of the quasiparticle wavefunction. There are two
equivalent theoretical approaches to calculate the supercurrent from the
model in Eq. (\ref{eq02}). The first approach is the scattering wave
function method \cite{ref11-BTK}, which provides a clear picture about
physical processes such as Andreev reflections and Andreev bound states. The
second approach is the Green's function method \cite{ref12-NEGF}, which
offers a unified formalism for general cases and allows effective
computational optimization to deal with Andreev bound states \cite{ref13-ABS}%
. However, applying these methods directly to a realistic Josephson junction
is not computationally affordable since the size of transport region is as
large as $100$ nm $\times $ $100$ nm $\times $ $1$ nm. The Fermi wavelength
of Al, on the other hand, is around $3.6$ \AA , estimated using the Fermi
energy $11.6$ eV. Consequently, the real space resolution should be around $%
1 $ \AA , resulting in a $10^{3}\times 10^{3}\times 10$ real space grid
which is too large for numerical simulations.

\begin{figure}
	\centering
	\includegraphics[width=8cm]{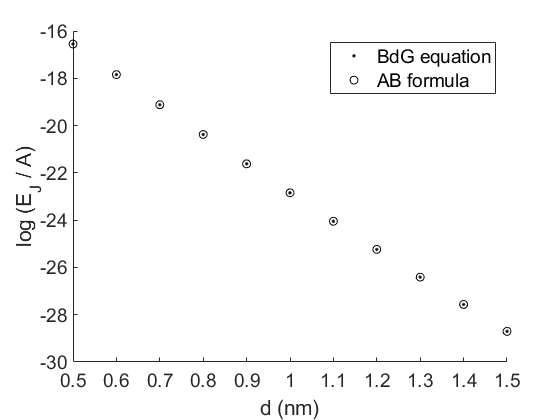}
	\caption{Josephson energy (log scale) as a function tunnel barrier width, calculated by solving the BdG equation and using the AB relation.
    }
	\label{fig01}
\end{figure}

To tackle the computational challenge, we adopt two approximations to
simplify the transport model. The first approximation is to use the
Ambegaokar--Baratoff (AB) relation 
\begin{equation}
I_{c}=\frac{\pi \Delta }{2eR_{N}},
\end{equation}%
which$\ $relates the critical current $I_{c}$ to the normal resistance $%
R_{N} $. The calculation of $R_{N}$ is much simpler than $I_{c}$ because it
only requires an evaluation of the transmission coefficient at the Fermi energy rather than an integration
over the entire energy range. Furthermore, the system size is reduced by
half since only the electron-like quasiparticles need to be considered. To
assess the accuracy of the AB relation for Josephson junctions, we compare
the critical current obtained by solving Eq. (\ref{eq02}) with that obtained from the AB relation for a Josephson junction with a square tunnel barrier of height $1.1$ eV. As shown in Fig. \ref{fig01}, the results of the two methods agree in high accuracy over a wide range of junction thickness, justifying the use of the AB relation in this study. Moreover, the exponential dependence of $E_{J}$ on the tunnel barrier thickness $d$ confirms that transport occurs in the tunneling regime.

The second approximation is to separate the transport direction from the
transverse directions. Because the junction operates in the tunneling
regime, transport is predominantly governed by the local barrier profile
along the transport direction, and transverse coupling is negligible. We
therefore partition the junction into many independent subsystems in the
transverse plane and approximate the total normal conductance as the sum of
the conductances of these subsystems. In the continuum limit, this sum
becomes an integral of the conductance density over the transverse
coordinates. Here, the conductance density is defined as the conductance per
unit area for a three-dimensional junction with a uniform cross section in
the transverse directions. This local thickness approximation is justified
by comparing the relevant length scales of the system. Four characteristic
lengths are involved: (1) The Fermi wave length in the Al leads%
\[
\lambda _{F}=\frac{2\pi }{k_{F}}=\frac{2\pi }{\sqrt{\frac{2mE_{F}}{\hbar ^{2}%
}}}\approx 3.6\text{ \AA ,} 
\]%
where\ $E_{F}=11.6$ eV the Fermi energy of Al bulk. (2) The wave function
decaying length in the tunnel barrier%
\[
\lambda _{D}=\frac{1}{\varkappa }=\frac{1}{\sqrt{\frac{2mU_{\text{barrier}}}{%
\hbar ^{2}}}}=1.9\text{ \AA ,} 
\]%
where $U_{\text{barrier}}=1.1$ eV is the effective tunnel barrier height of
the Al/AlO$_{\text{x}}$ interface. (3) The RMS of interface roughness along the
transport direction:\ $\sigma $ $\sim $ 0.1 nm. (4) The correlation length of
interface roughness in the transverse directions:\ $\xi \sim 10$ nm. The
change of the tunnel barrier thickness over a lateral distance $\lambda _{F}$
is estimated as $\frac{\lambda _{F}}{\xi }\sigma $. Since 
\[
\frac{\lambda _{F}}{\xi }\sigma \ll \lambda _{D}. 
\]%
the tunnel barrier can be viewed as an ensemble of locally uniform tunnel
barriers, which justifies the local thickness approximation.

\section{Results}

The interface roughness model and quantum transport framework are combined
to study the variability of the Josephson energy $E_{J}$ in Al/AlO$_{%
\text{x}}$/Al junctions. The junction is modeled as two free-electron
metallic leads separated by a three-dimensional tunnel barrier. The tunnel
barrier is characterized by a height $U\sim 1$ eV, a width $d\sim 1$ nm, and
a cross section $L_{x}\times L_{y}=200$ nm $\times $ $200$ nm, consistent
with Al/AlO$_{\text{x}}$ interface. The leads are characterized by a Fermi
energy $E_{F}=11.7$ eV and a superconducting gap $\Delta =0.2$ meV,
consistent with bulk Al. The interface roughness between the two leads and
the tunnel barrier is described by the Gaussian random field model in Eqs. (%
\ref{eq05},\ref{eq06}) with parameters RMS amplitude $\sigma $ and
transverse correlation length $\xi $. For simplicity, it is assumed that $\sigma $
and $\xi $ are identical for the two Al/AlO$_{\text{x}}$ interfaces in the Josephson junction. The resulting local barrier
thickness is obtained by adding the two interfacial height profiles to the
nominal thickness $d$, thus defining the spatially varying tunnel-barrier
profile used in the transport calculation.

\begin{figure}
	\centering
	\includegraphics[width=8cm]{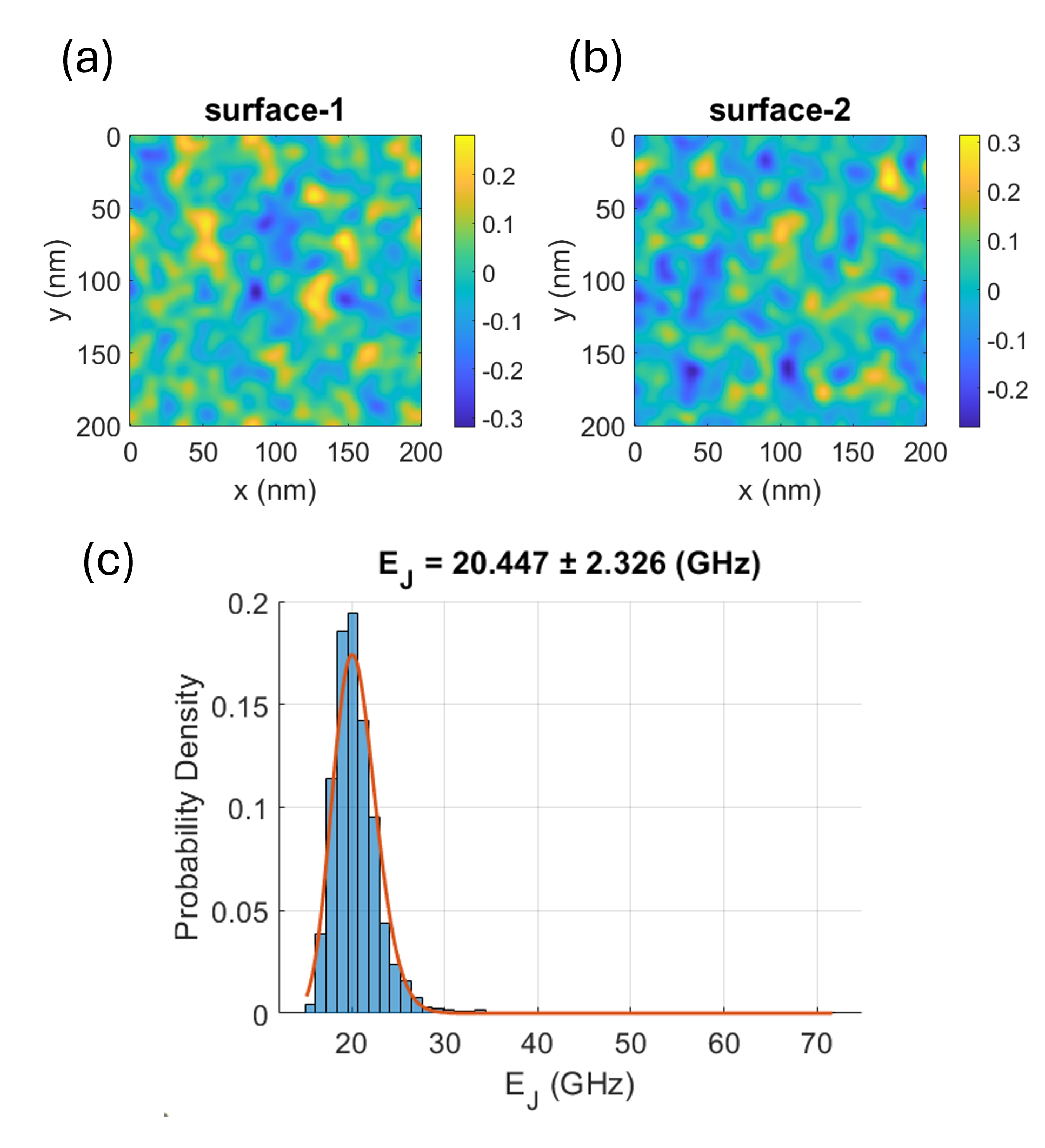}
	\caption{(a) and (b):\ Colormaps of Al/AlO$_{\text{x}}$ interface roughness. (c): Histogram of $E_{J}$ over $5,000$ Josephson junctions. The orange line is a fit to Eq. \ref{eq10}. 
    }
	\label{fig04}
\end{figure}

To be more specific, let us choose the tunnel barrier parameters
as $d=1.0$ nm and $U=1.1$ eV, and set the interface roughness parameters as 
$\sigma =0.85$ \AA\ and $\xi =10$ nm. Fig. \ref{fig04} (a)\ and (b) are the
colormaps of the resulting interface roughness, showing fluctuations at the 
scale of transverse correlation length $\xi $. Fig. \ref{fig04} (c) is the normalized histogram of the calculated $E_{J}$ over $5,000$ Josephson junctions with different tunnel barrier profiles. It is very interesting that the distribution of $E_{J}$ is not a Gaussian but skewed toward larger values. This feature can be understood as a result of the tunneling physics. Suppose the effective tunnel barrier width is normally distributed and since
$E_{J}$ has an exponential dependence on the tunnel barrier width, the distribution of $E_{J}$ must follow a log-normal distribution, and the corresponding probability density function is%
\begin{equation}
p\left( E_{J}>0\right) =\frac{1}{E_{J}\sigma _{J}\sqrt{2\pi }}\exp \left[ -%
\frac{\left( \log E_{J}-\mu _{J}\right) ^{2}}{2\sigma _{J}^{2}}\right] ,
\label{eq10}
\end{equation}%
where $\mu _{J}$ and $\sigma _{J}$ are the mean and standard deviation of $%
\log E_{J}$. Consequently, the mean and variance of $E_{J}$ are 
\[
\mathbb{E}\left[ E_{J}\right] =e^{\mu _{J}+\sigma _{J}^{2}/2}, 
\]%
and%
\[
\operatorname{Var}\left( E_{J}\right) =\left( e^{\sigma _{J}^{2}}-1\right) e^{2\mu
_{J}+\sigma _{J}^{2}}, 
\]%
respectively. These expressions show why the distribution of $E_{J}$ over the ensemble of the Josephson junctions becomes increasingly skewed and broadened as the underlying fluctuations grow. Fitting the histogram with Eq. (\ref{eq10}), we obtain%
\begin{equation}
E_{J}/h=20.447\pm 2.326\text{ GHz.}  \label{eq11}
\end{equation}%
For transmons, the qubit transition frequency is determined by $E_{C}$ and $%
E_{J}$ as\cite{ref02-qubit-design}
\[
\frac{\omega_{01}}{2\pi} = \frac{1}{h}(E_{1}-E_{0}) \approx \frac{1}{h} (\sqrt{%
8E_{C}E_{J}}-E_{C}). 
\]%
Therefore, the standard deviation of $\frac{\omega_{01}}{2\pi}$ is related to that of $%
E_{J}$ by 
\[
\Delta \frac{\omega_{01}}{2\pi} =\frac{1}{h}\sqrt{\frac{2E_{C}}{E_{J}}}\Delta E_{J}. 
\]%
Using the mean and standard deviation of $E_{J}$ in Eq. (\ref{eq11}) and
assuming $E_{C}/h=0.25$ GHz, we obtain the mean and standard deviation of $%
\omega_{01}/2\pi$ as%
\begin{equation}
\omega_{01}/2\pi=6.145\pm 0.364\text{ GHz.}  \label{eq12}
\end{equation}

Next, we investigate the dependence of the $E_{J}$ variability on the
roughness parameters $\sigma $ and $\xi $. Table \ref{tab01} lists the calculated mean and standard deviation of $E_{J}$ as functions of $\sigma $ and $\xi $,
while fixing the other device parameters at $U=1.1$ eV, $d=1.0$ nm, $%
\Delta =0.2$ meV, $E_{F}=11.7$ eV. Two main trends are observed. (1) With
increasing $\sigma $, both $\mathbb{E}\left[ E_{J}\right] $ and $\operatorname{Var}%
\left( E_{J}\right) $ increase. $\operatorname{Var}\left( E_{J}\right) $ increases
because the amplitude of thickness fluctuations scales with $\sigma $,
leading to stronger sample-to-sample variations in $E_{J}$. The rise in $%
\mathbb{E}\left[ E_{J}\right] $ reflects the exponential sensitivity of
tunneling to the local barrier thickness: the total conductance is dominated
by rare, anomalously thin regions, which become thinner and more influential
as $\sigma $ grows. (2) With increasing $\xi $, $\mathbb{E}\left[ E_{J}%
\right] $ decreases slightly, whereas $\operatorname{Var}\left( E_{J}\right) $
increases markedly, and both trends saturate once $\xi $ becomes comparable
to the lateral size of the junction. The increase of $\operatorname{Var}\left(
E_{J}\right) $ can be understood as a consequence of reduced self-averaging.
The junction may be viewed as comprising many parallel tunneling channels,
each characterized by a correlation area of order $\xi ^{2}$. A larger $\xi $
implies fewer statistically independent channels, which in turn leads to
enhanced sample-to-sample fluctuations and a larger variance. Meanwhile, the
slight decrease of $\mathbb{E}\left[ E_{J}\right] $ is associated with
finite-size effect. For a junction with a finite cross section, increasing $%
\xi $ reduces the number of independent segments across the interface.
Consequently, the probability of realizing rare, exceptionally thin barrier
regions that would otherwise dominate the tunneling contribution is reduced,
resulting in a modest suppression of $\mathbb{E}\left[ E_{J}\right] $. When $%
\xi $ approaches the junction dimension, the effective number of independent
segments becomes $O\left( 1\right) $, and both the mean and variance of $%
E_{J}$ become insensitive to further increases in $\xi $, resulting in
saturation behavior.

\begin{table}[t]
\caption{Josephson energy $E_{J}/h$ (GHz) as a function of RMS roughness $%
\protect\sigma $ and transverse correlation length $\protect\xi $. Values
are given as mean $\pm $ standard deviation.}\centering \setlength{%
\tabcolsep}{10pt} \renewcommand{\arraystretch}{1.2} 
\begin{tabular}{cccc}
\hline
$\xi$ (nm) & $\sigma = 0.08$ nm & $\sigma = 0.09$ nm & $\sigma = 0.10$ nm \\ 
\hline
10 & $17.92 \pm 1.70$ & $23.57 \pm 3.20$ & $32.32 \pm 6.15$ \\ 
20 & $17.86 \pm 3.05$ & $23.39 \pm 5.45$ & $31.81 \pm 9.82$ \\ 
30 & $17.72 \pm 3.99$ & $23.05 \pm 6.87$ & $31.06 \pm 11.94$ \\ 
40 & $17.54 \pm 4.70$ & $22.67 \pm 7.89$ & $30.27 \pm 13.35$ \\ 
50 & $17.37 \pm 5.24$ & $22.32 \pm 8.64$ & $29.56 \pm 14.35$ \\ \hline
\end{tabular}%
\label{tab01}
\end{table}

\section{Summary}

To summarize, we have investigated the variability of the Josephson energy $%
E_{J}$ in Al/AlO$_{\text{x}}$/Al junctions arising from interface roughness.
The roughness is modeled using a Gaussian random field characterized by the
RMS amplitude $\sigma $ and the transverse correlation length $\xi $. The
values of these parameters are taken from the experimental literature and extracted from our atomistic MD simulations. The quantum transport model is 
based on the Ambegaokar--Baratoff relation combined with a local thickness approximation, both of which are justified and numerically verified for the Josephson junctions. Numerical simulations over an ensemble of $5,000$ samples show that $E_{J}$ follows a log-normal distribution. The mean value of $E_{J}$ increases with $\sigma $ and decreases slightly with $\xi $, while the variance of $E_{J}$ increases with both $\sigma $ and $\xi $. In this work, we have assumed that $\sigma $ and $\xi $ are independent structural parameters and are identical for the two interfaces in Al/AlO$_{\text{x}}$/Al junctions. While reasonable, these assumptions require careful experimental verification under different growth and process conditions. Connecting our model to the experimental characterization of the fabricated Josephson junctions and combining the superconductivity solvers employed here with computational microwave engineering software\cite{qtcad} will be important directions for future work, opening the door to the co-design of Josephson junctions and their neighboring superconducting-circuit elements.

\section{Acknowledgements}

We thank the Digital Research Alliance of Canada for the computational facilities that made this work possible. H.G. is grateful to NSERC of Canada for partial financial support. The authors thank Drs. Rapha\"{e}l Prentki, Pericles Philippopoulos, and Igor Benek-Lins for useful discussions regarding the interface roughness model, the transmon and the superconducting qubits. 
\nocite{*}
\bibliography{references}

@article{ref01-outlook,
	author = {Devoret, Michel H. and Schoelkopf, Robert J.},
	title = {Superconducting circuits for quantum information: an outlook},
	journal = {Science},
	year = {2013},
	volume = {339},
	number = {6124},
	pages = {1169--1174},
	date = {2013-03-08},
	doi = {10.1126/science.1231930},
	url = {https://pubmed.ncbi.nlm.nih.gov/23471399/},
	urldate = {2026-01-07}
}

@article{ref02-qubit-design,
	author = {Koch, Jens and Yu, Terri M. and Gambetta, Jay and Houck, A. A. and Schuster, D. I. and Majer, J. and Blais, Alexandre and Devoret, M. H. and Girvin, S. M. and Schoelkopf, R. J.},
	title = {Charge-insensitive qubit design derived from the Cooper pair box},
	journal = {Physical Review A},
	year = {2007},
	volume = {76},
	pages = {042319},
	doi = {10.1103/PhysRevA.76.042319},
	url = {https://link.aps.org/doi/10.1103/PhysRevA.76.042319},
	urldate = {2026-01-07}
}

@misc{ref03-Condor,
	author = {Gambetta, Jay},
	title = {The hardware and software for the era of quantum utility is here},
	organization = {IBM Quantum},
	year = {2023},
	date = {2023-12-04},
	url = {https://www.ibm.com/quantum/blog/quantum-roadmap-2033},
	urldate = {2026-01-07}
}

@article{ref04-RSA2048,
	author = {Gidney, Craig and Eker{\aa}, Martin},
	title = {How to factor 2048 bit {RSA} integers in 8 hours using 20 million noisy qubits},
	journal = {Quantum},
	year = {2021},
	volume = {5},
	pages = {433},
	doi = {10.22331/q-2021-04-15-433},
	eprint = {1905.09749},
	eprinttype = {arxiv},
	eprintclass = {quant-ph},
	url = {https://quantum-journal.org/papers/q-2021-04-15-433/},
	urldate = {2026-01-07}
}

@article{ref05-WaferScaleJJ,
	author = {Kreikebaum, John Mark and O'Brien, K. P. and Morvan, A. and Siddiqi, Irfan},
	title = {Improving wafer-scale {Josephson} junction resistance variation in superconducting quantum coherent circuits},
	journal = {Superconductor Science and Technology},
	year = {2020},
	volume = {33},
	number = {6},
	pages = {06LT02},
	doi = {10.1088/1361-6668/ab8617},
	eprint = {1909.09165},
	eprinttype = {arxiv},
	eprintclass = {quant-ph},
	url = {https://arxiv.org/abs/1909.09165},
	urldate = {2026-01-07}
}

@article{ref06-JJ-reprod1,
	author = {Pishchimova, Anastasiya A. and Smirnov, Nikita S. and Ezenkova, Daria A. and Krivko, Elizaveta A. and Zikiy, Evgeniy V. and Moskalev, Dmitry O. and Ivanov, Anton I. and Korshakov, Nikita D. and Rodionov, Ilya A.},
	title = {Improving Josephson junction reproducibility for superconducting quantum circuits: junction area fluctuation},
	journal = {Scientific Reports},
	year = {2023},
	date = {2023-04-25},
	volume = {13},
	number = {1},
	eid = {6772},
	doi = {10.1038/s41598-023-34051-9},
	pmid = {37185459},
	pmcid = {PMC10130087},
	url = {https://pmc.ncbi.nlm.nih.gov/articles/PMC10130087/},
	urldate = {2026-01-07}
}

@article{ref07-JJ-reprod2,
	author = {Moskalev, Dmitry O. and Zikiy, Evgeniy V. and Pishchimova, Anastasiya A. and Ezenkova, Daria A. and Smirnov, Nikita S. and Ivanov, Anton I. and Korshakov, Nikita D. and Rodionov, Ilya A.},
	title = {Optimization of shadow evaporation and oxidation for reproducible quantum Josephson junction circuits},
	journal = {Scientific Reports},
	year = {2023},
	date = {2023-03-13},
	volume = {13},
	number = {1},
	eid = {4174},
	doi = {10.1038/s41598-023-31003-1},
	url = {https://www.nature.com/articles/s41598-023-31003-1},
	urldate = {2026-01-07}
}

@article{ref08-mitigation,
	author = {Song, S. and Sun, Y. and Xu, J. and Han, Z. and Yang, X. and Wang, X. and Li, S. and Lan, D. and Zhao, J. and Tan, X. and Yu, Y.},
	title = {Mitigation of critical current fluctuation of Josephson junctions in superconducting quantum circuits},
	journal = {Applied Physics Letters},
	year = {2021},
	volume = {118},
	number = {24},
	eid = {244004},
	doi = {10.1063/5.0049637},
	url = {https://pubs.aip.org/aip/apl/article-abstract/118/24/244004/238962/Mitigation-of-critical-current-fluctuation-of},
	urldate = {2026-01-07}
}

@article{ref11-BTK,
	author = {Blonder, G. E. and Tinkham, M. and Klapwijk, T. M.},
	title = {Transition from metallic to tunneling regimes in superconducting microconstrictions: Excess current, charge imbalance, and supercurrent conversion},
	journal = {Physical Review B},
	year = {1982},
	volume = {25},
	number = {7},
	pages = {4515--4532},
	doi = {10.1103/PhysRevB.25.4515}
}

@article{ref12-NEGF,
	author = {Sun, Q. F. and Wang, B. and Wang, J. and Lin, T. H.},
	title = {Electron transport through a mesoscopic hybrid multiterminal resonant-tunneling system},
	journal = {Physical Review B},
	year = {2000},
	volume = {61},
	number = {7},
	pages = {4754--4761},
	doi = {10.1103/PhysRevB.61.4754}
}

@article{ref13-ABS,
	author = {Zhu, Yu and Sun, Qing{-}feng and Lin, Tsung{-}han},
	title = {Probing spin states of coupled quantum dots by a dc Josephson current},
	journal = {Physical Review B},
	year = {2002},
	volume = {66},
	number = {8},
	pages = {085306},
	date = {2002-08-09},
	doi = {10.1103/PhysRevB.66.085306},
	url = {https://link.aps.org/doi/10.1103/PhysRevB.66.085306},
	urldate = {2026-01-07}
}

@article{ref22-Al02,
  author       = {Nik, Samira and Krantz, Philip and Zeng, Lunjie and Greibe, Tine and Pettersson, Henrik and Gustafsson, Stefan and Delsing, Per and Olsson, Eva},
  title        = {Correlation between {Al} grain size, grain boundary grooves and local variations in oxide barrier thickness of {Al/AlO$_x$/Al} tunnel junctions by transmission electron microscopy},
  journal      = {SpringerPlus},
  date         = {2016-07-13},
  year         = {2016},
  volume       = {5},
  number       = {1},
  pages        = {1067},
  doi          = {10.1186/s40064-016-2418-8},
  url          = {https://link.springer.com/article/10.1186/s40064-016-2418-8},
  urldate      = {2026-01-27},
  pmid         = {27462515},
  pmcid        = {PMC4943912},
}

@article{ref23-Al03,
  author       = {Zeng, Lunjie and Tran, Dung Trung and Tai, Cheuk-Wai and Svensson, Gunnar and Olsson, Eva},
  title        = {Atomic structure and oxygen deficiency of the ultrathin aluminium oxide barrier in {Al/AlO$_x$/Al} {Josephson} junctions},
  journal      = {Scientific Reports},
  date         = {2016-07-12},
  year         = {2016},
  volume       = {6},
  pages        = {29679},
  doi          = {10.1038/srep29679},
  url          = {https://www.nature.com/articles/srep29679},
  urldate      = {2026-01-27},
}

@article{Zeng2015,
	author = {Zeng, L. J. and Nik, S. and Greibe, T. and Wilson, C. M. and Delsing, P. and Olsson, E.},
	title = {Direct observation of the thickness distribution of ultra thin {AlO$_x$} barrier in {Al/AlO$_x$/Al} {Josephson} junctions},
	journal = {Journal of Physics D: Applied Physics},
	volume = {48},
	number = {39},
	pages = {395308},
	year = {2015},
	doi = {10.1088/0022-3727/48/39/395308},
	url = {https://doi.org/10.1088/0022-3727/48/39/395308},
	eprint = {1407.0173},
	eprinttype = {arxiv}
}

@article{Fritz2019,
	author = {Fritz, S. and Radtke, L. and Schneider, R. and Weides, M. and Gerthsen, D.},
	title = {Optimization of {Al/AlO$_x$/Al}-layer systems for {Josephson} junctions from a microstructure point of view},
	journal = {Journal of Applied Physics},
	volume = {125},
	number = {16},
	pages = {165301},
	year = {2019},
	doi = {10.1063/1.5089871},
	url = {https://doi.org/10.1063/1.5089871},
	eprint = {1901.09625},
	eprinttype = {arxiv}
}

@article{ref24-chgnet,
	author = {Deng, Bowen and Zhong, Peichen and Jun, KyuJung and Riebesell, Janosh and Han, Kevin and Bartel, Christopher J. and Ceder, Gerbrand},
	title = {CHGNet as a pretrained universal neural network potential for charge-informed atomistic modelling},
	journal = {Nature Machine Intelligence},
	year = {2023},
	volume = {5},
	pages = {1031--1041},
	doi = {10.1038/s42256-023-00716-3},
}

@Article{pappas2024alternating,
  author    = {Pappas, David P and Field, Mark and Kopas, Cameron J and Howard, Joel A and Wang, Xiqiao and Lachman, Ella and Oh, Jinsu and Zhou, Lin and Gold, Alysson and Stiehl, Gregory M and others},
  title     = {Alternating-bias assisted annealing of amorphous oxide tunnel junctions},
  number    = {1},
  pages     = {150},
  volume    = {5},
  journal   = {Communications Materials},
  publisher = {Nature Publishing Group UK London},
  year      = {2024},
}

@misc{qtcad,
  author       = {{Nanoacademic Technologies Inc.}},
  title        = {{QTCAD: Superconductors -- Theory}},
  year         = {2026},
  howpublished = {\url{https://docs.nanoacademic.com/qtcad/theory_sc/theory_sc/}},
}

@article{ref31_Liu_2023,
	title = {Unveiling atomic structure and chemical composition of the Al/AlOx/Al Josephson junctions in qubits},
	journal = {Applied Surface Science},
	volume = {640},
	pages = {158337},
	year = {2023},
	issn = {0169-4332},
	doi = {https://doi.org/10.1016/j.apsusc.2023.158337},
	url = {https://www.sciencedirect.com/science/article/pii/S0169433223020172},
	author = {Xiaotao Liu and Kejia Pan and Zhen Zhang and Zhiyuan Feng},
}

\end{document}